# Structural and electronic properties of highly doped topological insulator Bi$_2$Se$_3$ crystals

Helin Cao [*,1,2], Suyang Xu[3, 4], Ireneusz Miotkowski[1], Jifa Tian[1, 2], Deepak Pandey[1,2], M. Zahid Hasan[3, 4], Yong P. Chen[*,1, 2, 5]

[1] Department of Physics, Purdue University, 525 Northwestern Ave., West Lafayette, IN 47907, USA
[2] Birck Nanotechnology Center, Purdue University, 1205 W. State St., West Lafayette, IN 47907, USA
[3] Joseph Henry Laboratories, Department of Physics, Princeton University, Jadwin Hall, Princeton, New Jersey 08544, USA
[4] Princeton Institute for Science and Technology of Materials, Princeton University, 70 Prospect Ave., Princeton, New Jersey 08544, USA
[5] School of Electrical and Computer Engineering, Purdue University, 465 Northwestern Ave., West Lafayette, IN 47907 USA

**Abstract**   We present a study of the structural and electronic properties of highly doped topological insulator Bi$_2$Se$_3$ single crystals synthesized by the Bridgeman method. Lattice structural characterizations by X-ray diffraction, scanning tunneling microscopy, and Raman spectroscopy confirmed the high quality of the as-grown single crystals. The topological surface states in the electronic band structure were directly revealed by angle-resolved photoemission spectroscopy. Transport measurements showed the conduction was dominated by the bulk carriers and confirmed a previously observed bulk quantum Hall effect in such highly doped Bi$_2$Se$_3$ samples. We briefly discuss several possible strategies of reducing bulk conductance.

**Introduction** The electronic properties of Bi$_2$Se$_3$, a semiconductor made of van der Waals coupled stacking "quintuple layers" (QL), were thought to be well understood since it had been studied for decades in light of its excellent thermoelectric properties. However, recently, it has been revealed that Bi$_2$Se$_3$ belongs to a new class of quantum materials: 3D topological insulators (TI)[1,2]. In a 3D TI, the bulk has a band gap (~0.3 eV for Bi$_2$Se$_3$), and the surface has non-trivial topologically protected surface states (SS)[3-7], which give rise to 2D Dirac fermions with an odd number of Dirac cones (1 for Bi$_2$Se$_3$) and spin-momentum locking. Most as-grown Bi$_2$Se$_3$ samples have significant amount of (uncontrolled) Se vacancies that cause unintentional n-type bulk doping. Previous, well-documented magnetotransport studies[8-10] of Bi$_2$Se$_3$ with bulk carrier densities between ~ $10^{17}$ and $10^{19}$ cm$^{-3}$ show standard 3D transport behavior of a doped bulk semiconductor. Interestingly, recent measurements on Bi$_2$Se$_3$ in lower and higher carrier density regimes have both revealed novel magnetotransport behaviors. In very low bulk doping (with carrier density <$10^{17}$ cm$^{-3}$) samples (synthesized with compensating dopants, such as Sb), quantum

---

oscillations of the 2D Dirac fermions attributed to TI SS have been observed[11]. In very high bulk doping (carrier density $>\sim 10^{19}$ cm$^{-3}$) samples, a bulk quantum Hall effect (QHE) with a 2D like transport behavior arising from parallel QLs in the 3D bulk has been observed[12]. Here, we present comprehensive structural and electronic characterizations on such highly doped $Bi_2Se_3$. The structure of the crystals was analyzed by X-ray diffraction (XRD), scanning tunneling microscopy (STM), and Raman spectroscopy. The TI SS in our sample was directly revealed by angle-resolved photoemission spectroscopy (ARPES). On the other hand, our transport measurements show that the electronic conduction is dominated by the bulk. We present the temperature (T) dependent resistance data on samples with different thicknesses (from ~ 60 nm to 310 nm), and magnetotransport data on one of the sample ("A") confirming the previously observed bulk QHE[12] (up to higher magnetic field than that in Ref. 12). Our results underscore the challenge of accessing surface state electronic transport (due to high bulk conduction) in as-grown $Bi_2Se_3$ bulk crystals. At the end of this paper, we briefly discuss several possible strategies of reducing bulk doping or conduction.

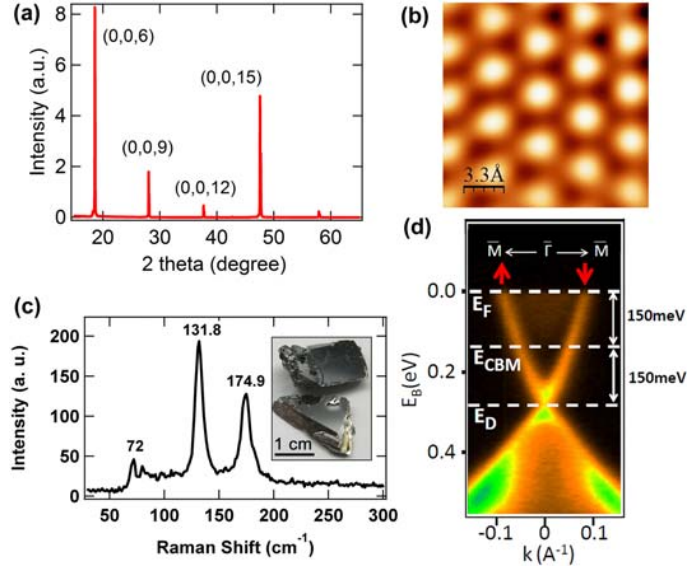

**Figure 1** (a) Representative XRD pattern on single crystal $Bi_2Se_3$ (001) surface. The peaks are labeled with (*hkl*) Miller indices. (b) Atomic resolution STM image of a cleaved surface of $Bi_2Se_3$ crystal. (c) Representative Raman spectrum measured on a $Bi_2Se_3$ crystal. Three characteristic Raman peaks are labeled. Inset: optical image of the $Bi_2Se_3$ single crystals. (d) High-resolution ARPES energy-momentum dispersion band mapping along a pair of time-reversal invariant points M-Γ-M on $Bi_2Se_3$.

**Results and discussion** High quality $Bi_2Se_3$ single crystals (optical images shown in the inset of Fig. 1c) were synthesized by the Bridgman technique (see Supplementary Information). Fig. 1a shows an XRD pattern measured on $Bi_2Se_3$ (001) surface. The XRD reflections are attributed to $Bi_2Se_3$ (0,0,6) (0,0,9) (0,0,12) (0,0,15) planes, confirming the sample's single crystalline structure. From powder XRD data (not shown), we extract two lattice constants, a = 4.137 Å and c = 28.679 Å, which are consistent with the previously reported values[13,14]. The atomic resolution STM image (Fig. 1b) of a cleaved $Bi_2Se_3$ (001) surface was measured in ultrahigh vacuum, showing a lattice constant a = 4.2 Å, confirming the XRD result. The lattice structure of our samples was further investigated by Raman spectroscopy. A representative Raman spectrum is shown in Fig. 1c. Three Raman peaks at proximately 72 cm$^{-1}$, 131.8 cm$^{-1}$, and 174.9 cm$^{-1}$ agree well with the characteristic lattice vibration modes $A_{1g}^1$, $E_g^2$ and $A_{1g}^2$ observed in previous study[15]. The topological SS of our samples were directly revealed by ARPES as shown in Fig. 1d. The Fermi level is located at 150meV above the bulk conduction band minimum (CBM) in this sample. The Fermi wave vector and Fermi velocity of the SS are found to be $k_F = 0.1$ Å$^{-1}$ and $v_F = 5\times10^5$ m/s, similar to previous APRES measurements[1] on $Bi_2Se_3$ (see Supplementary Information for more analysis on the ARPES data).

Fig. 2a shows the 4-terminal longitudinal resistance ($R_{xx}$) of a 150 nm-thick exfoliated flake (sample A) measured from room T down to 5K, displaying a metallic behavior, as qualitatively expected for the highly doped bulk. The T dependence of $R_{xx}$ can be fitted to a simplified phenomenological model developed for doped $Bi_2Te_3$ bulk crystals[16] (which shows generally similar transport properties as doped $Bi_2Se_3$)

$$R_{xx} = R_0 + \alpha \times e^{-\theta/T} + \beta \times T^2 \qquad (1)$$

$R_0$, a low T residual resistance, corresponds to the contributions of impurity scattering. Phonon scattering and electron-electron scattering give rise to the exponential and the quadratic terms respectively. We found $R_0 = 22.46$ Ω, $\alpha = 13.5$ Ω, $\theta = 217$ K and $\beta = 0.00009$ Ω/K$^2$ give the best fit (as shown by the red line in Fig. 2a) to the experimental data (round black circle in Fig. 2a). The T dependence is dominated by the phonon scattering[16] and the fitting parameter $\theta$ corresponds to an effective phonon frequency $\omega = k_B\theta/\hbar = 3.1\times10^{13}$ rad/s. The very small value of $\beta$ indicates that e-e scattering effect is negligible in our sample. We measured T dependence resistivity on 6 samples with different thicknesses. As an indicator of the metallic behavior, we take the high T (270 K) resistance normalized by the respective low T (15 K) value, and plot the ratio against the sample thickness (Fig. 2b). Interestingly, the thinner samples appear to be "less metallic" as measured by this ratio. Fig. 2b suggests that thinning down the crystal thickness can be an effective means to reduce the metallic bulk conduction of $Bi_2Se_3$ (even for samples with high bulk doping) that may help bring out the SS transport signatures that are often overwhelmed by the bulk conduction.

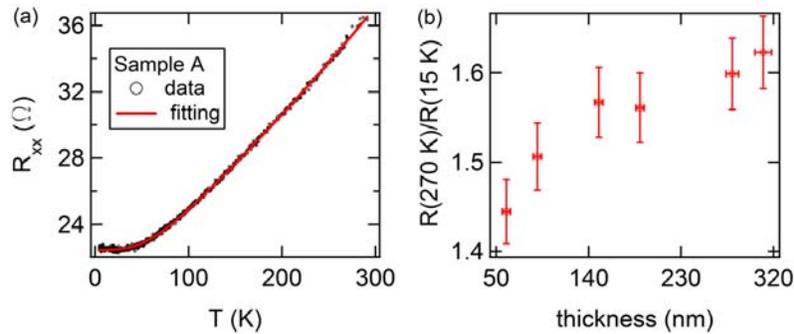

**Figure 2** (a) Temperature dependence of four-terminal longitudinal resistance ($R_{xx}$) in sample A. The red line shows the fitting to Eq. 1. (b) R(T = 270 K)/R(T = 15 K) plotted against sample thicknesses.

Fig. 3a shows $R_{xx}$ and Hall resistance $R_{xy}$ for sample A as functions of perpendicular magnetic field (B) applied along the c-axis at 340 mK. The carrier (n-type) density and mobility extracted from the low B measurements are $4.7\times10^{19}$ cm$^{-3}$ and ~400 cm$^2$/Vs respectively. At higher B, $R_{xx}$ oscillates periodically in 1/B (with a period $B_F$), which can be interpreted as SdH oscillations due to the formation of Landau levels (LL). The N$^{th}$ minimum of $R_{xx}$, counting from B = $B_F$ (which defines N=1), corresponds to the N$^{th}$ LL (labeled in Fig. 3a). We plot the assigned LL index N against the inverse of the magnetic field (B) positions of the observed minima in $R_{xx}$(B) in the inset of Fig 3a. The black solid line is a linear fit with N-axis intercept 0±0.02 and slope $B_F$ = 163T, corresponding to a bulk carrier Fermi wave vector $k_F$ = 0.07 Å$^{-1}$. Furthermore, accompanying the minima in $R_{xx}$, $R_{xy}$ shows developing quantized plateaus. In Fig. 3b, we plot normalized quantized Hall step size $\Delta(1/R_{xy})$ and $\Delta R_{xy}$ (difference between two adjacent plateaus in $1/R_{xy}$ and $R_{xy}$ respectively, then normalized by their own values at N = 6) as functions of LL index N. It clearly shows $\Delta(1/R_{xy})$ is largely independent with increasing LL index, while $\Delta R_{xy}$ decreases. The approximately constant value of $\Delta(1/R_{xy})$ for different LLs is ~1.2$e^2$/h per QL (the number of QLs is determined by sample's thickness, where the scaling of $\Delta(1/R_{xy})$ with the thickness as observed previously[12] further confirms the transport is dominated by the bulk). The quantization in $R_{xy}$ can be interpreted as a "bulk QHE"[12] attributed to parallel 2D electron gas's arising from the stacking QLs in highly doped Bi$_2$Se$_3$ crystals (carrier density > ~$10^{19}$ cm$^{-3}$, where the Se vacancies may help reduce the electronic couplings between the QLs), and not caused by SS or any other surface conduction channels. The quantized Hall step size in $1/R_{xy}$ was examined down to lower LLs compared to our previous study[12], and confirmed to remain approximately constant for different LLs. Our results (here and in Ref. 12), along with those from earlier experiments[10, 11, 17-19] on less-doped Bi$_2$Se$_3$, demonstrate the rich physics in the magnetotransport of Bi$_2$Se$_3$ in different regimes of bulk carrier densities.

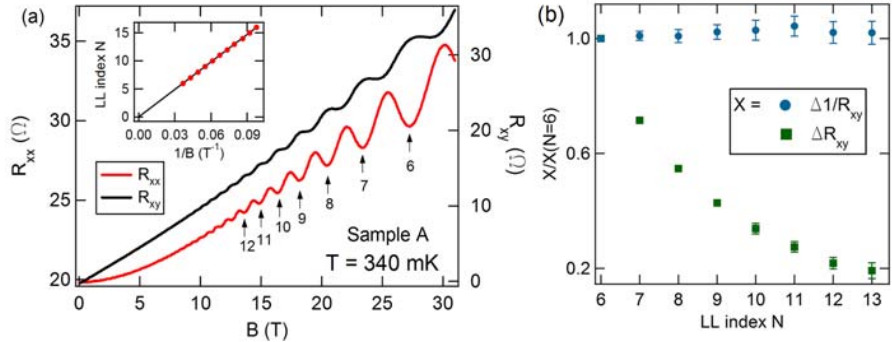

**Figure 3** (a) Hall resistance ($R_{xy}$) and $R_{xx}$ of sample A as functions of magnetic field (B, perpendicular to the QLs) measured at T = 340 mK. The minima in $R_{xx}$ corresponding to a series of Landau Level indices are labeled with arrows. The inset: LL fan diagram. (b) Normalized step size in $1/R_{xy}$ and $R_{xy}$ plotted against LL index (N).

**Conclusion** High quality Bi$_2$Se$_3$ single crystals were synthesized by Bridgman techniques. The lattice structure is characterized by XRD, STM and Raman spectroscopy that confirmed the excellent crystalline quality. Our as-grown crystals are found to have high bulk carrier density (>$10^{19}$ cm$^{-3}$, presumably due to substantial n-type doping by Se vacancies). The temperature dependence of resistance confirms the metallic behavior, which is weaker for thinner samples. We conducted magnetotransport measurements in a higher B field to reach lower LL compared to our previous study[12] and confirmed the bulk QHE previously observed. Accessing SS transport in as-grown $Bi_2Se_3$ remains challenging. There are several possible strategies to reduce the bulk conduction: for example, 1) thinning down the thickness of the crystal[20,21]; 2) adding more Se during growth to reduce Se vacancies (see Fig. S1 in Supplementary Information); 3) growing mixed crystals such as $Bi_2Te_2Se$ which has been shown to have a large bulk resistivity (low bulk carrier density)[22, 23]; 4) growing crystals with compensating dopants (e.g., Sb[11] or Ca[24]) to reduce bulk doping. Employing one or more of such strategies will likely be important to prepare TI materials for transport studies and device applications of TI SS.

**Acknowledgements** We acknowledge support from DARPA MESO program (Grant N66001-11-1-4107). STM measurements and part of the transport measurements were carried out at Argonne National Laboratory Center for Nanoscale Materials (CNM) under the auspices of CNM user research program (CNM-998). Part of the magnetotransport measurements were performed at the National High Magnetic Field Laboratory (NHMFL). The Princeton-led synchrotron x-ray-based measurements are supported by the Office of Basic Energy Sciences, U.S. Department of Energy (grants DE-FG-02-05ER46200, and AC03-76SF00098). M.Z.H. acknowledges visiting-scientist support from Lawrence Berkeley National Laboratory (LBNL) and additional support from the A. P. Sloan Foundation. The ARPES measurements using synchrotron X-ray facilities are supported by the Advanced Light Source in the LBNL. We thank E. Palm (NHMFL), L. Engel (NHMFL), J. J. Jaroszynski (NHMFL), N. P. Guisinger (ANL), B. Fisher (ANL) and P. Metcalf (Purdue) for experimental assistance.

**Supplementary Information**

Sample growth

High quality $Bi_2Se_3$ single crystals were synthesized by the Bridgman technique using a two-step procedure. In the first step, the raw material was synthesized from high purity elements in a two-zone horizontal furnace with independent temperature control. All the starting materials, the 5N (99.999% pure) Bi and 5N Se, were deoxidized and purified further by multiple vacuum distillations under dynamic vacuum of $10^{-7}$ torr. The synthesis was made in vitreous carbon boats to avoid possible contamination from quartz. After the synthesis was completed at $900^{\circ}$ C, the pre-reacted charge was slowly cooled down under a controlled pressure of Se. The charge was then transferred into a carbonized quartz ampoule. The growth ampoule was placed in a vertical Bridgman three-zone furnace with independent temperature control. The linear gradient in the growth zone was set to $5^{\circ}$ C/cm. The radial gradient inside the growth zone was symmetric and estimated to be less than $0.5^{\circ}$ C/cm. The typical speed for moving the ampoule through the temperature gradient was between 0.5 to 1.5 mm/hour.

Measurements setup

As-grown $Bi_2Se_3$ crystals were cleaved into smaller pieces for various measurements. X-ray diffraction (XRD) was measured by a Bragg-Brentano geometry Bruker D8 Focus X-ray diffractometer with Cu Kα radiation. For electronic transport measurements, thin flakes exfoliated from as-grown crystals were fabricated into devices with quasi-Hall-bar geometry. The exfoliated surface (001) is parallel to quintuple layers and perpendicular to the c-axis. Raman spectra were measured in ambient using a 532 nm excitation laser with circular polarization and ~200 μW incident power. The ARPES measurements were performed at the Advanced Light Source in Lawrence Berkeley National Laboratory.

ARPES measurement of topological surface states (SS)

The topological SS of our samples were directly revealed by angle-resolved photoemission spectroscopy (ARPES). The typical energy and momentum resolution was 15 meV and 1% of surface Brillouin Zone (BZ). The samples were cleaved *in situ* at low T (10 to 20K) and chamber pressure less than $5\times10^{-11}$ Torr, resulting in shiny flat surfaces. Fig. 1d in the main text shows an energy-momentum dispersion band map of our sample. The V-shaped topological surface state band is clearly revealed, confirming the TI nature of our $Bi_2Se_3$ crystal. The $k_F$ width of the Dirac surface band ($\delta k = 0.015$ Å$^{-1}$) indicates an ARPES-measured mean free path $L = 1/\delta k = 67$ nm for the $Bi_2Se_3$ SS. We note the ARPES data show a spectral weight suppression (resulting in a gap-like feature) at the Dirac point of the topological surface state. Such a feature has been found to occur quite often in many topological insulator samples (without explicit time-reversal symmetry breaking mechanisms such as magnetic doping), and can be attributed to various extrinsic factors, such as top-layers-relaxations, sample inhomogeneity (spatial energy-momentum fluctuation of the surface states), and surface chemical imperfections etc. (see Ref. S1 and Supplementary Information of Ref. S2).

Effects of adding more Se during growth

We have synthesized $Bi_2Se_3$ crystals by mixing Bi and Se with a non-stoichiometric ratio of 2:4.2, and found significantly reduced bulk carrier densities (by more than 1 order of magnitude) in these samples compared with nominally-stoichiometric samples, as shown in Fig. S1.

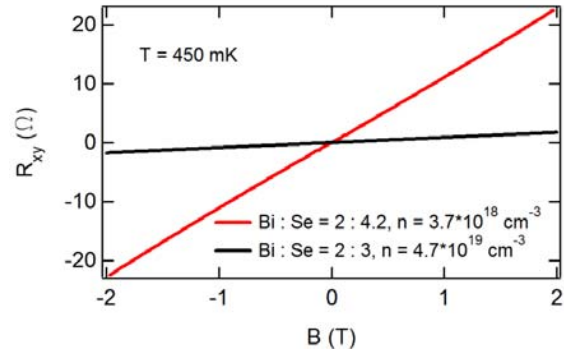

**Figure S1** Hall resistance of two samples grown with two different ratios of elements (Bi/Se = 2/3 and 2/4.2). The thicknesses of both samples (exfoliated from the as-grown crystals) are ~ 150 nm.

**Reference**
[S1] S. Xu *et al.*, arXiv:1206.0278 v2 (2012).
[S2] S. Xu *et al.*, Nature Phys. **8**, 616 (2012).